# Geometrodynamics of Spinning Light

Konstantin Y. Bliokh[1,2,3], Avi Niv[1], Vladimir Kleiner[1], and Erez Hasman[1]

[1]*Micro and Nanooptics Laboratory, Faculty of Mechanical Engineering and Russell Berrie Nanotechnology Institute, Technion–Israel Institute of Technology, Haifa 32000, Israel, http://www.technion.ac.il/optics/*

[2]*Institute of Radio Astronomy, 4 Krasnoznamyonnaya St., Kharkov 61002, Ukraine*

[3]*Nonlinear Physics Centre, Research School of Physical Sciences and Engineering, Australian National University, Canberra ACT 0200, Australia*

**The semiclassical evolution of spinning particles has recently been re-examined in condensed matter physics, high energy physics, and optics, resulting in the prediction of the intrinsic spin Hall effect associated with the Berry phase. A fundamental nature of this effect is related to the spin-orbit interaction and topological monopoles. Here we report a unified theory and a direct observation of two mutual phenomena: a spin-dependent deflection (the spin Hall effect) of photons and the precession of the Stokes vector along the coiled ray trajectory of classical geometrical optics. Our measurements are in perfect agreement with theoretical predictions, thereby verifying the dynamical action of the topological Berry-phase monopole in the evolution of light. These results may have promising applications in nano-optics and can be immediately extrapolated to the evolution of massless particles in a variety of physical systems.**

The discovery of the geometric Berry's phases in the 1980s raised interest in universal geometrical structures, such as topological monopoles underlying the evolution of quantum particles[1,2]. The topological monopoles appear in the points of level degeneracy in parameter space, producing the Berry curvature responsible for the parallel transport of the particle state vector. In the 1990s, it was shown that the Berry phase isnot a purely geometrical phenomenon, but also a dynamical effect. As a result, the semiclassical equations of motion have been re-examined, where the Berry curvature occurs as an external field affecting the motion of the particle[3,4].

As applied to the evolution of particles with a spin, this has led to the explanation of the anomalous Hall effect[5] and the prediction of the intrinsic spin Hall effect (SHE)[6,7] in semiconductor systems. For relativistic spinning particles, the Berry phase and the SHE are two manifestations of the spin-orbit interaction[8,9], which describes the mutual influence of the spin (polarization) and trajectory of the particle. In the massless case, this is associated with a topological monopole that appears in the Dirac point, i.e., at the origin of momentum space[5,6,9,10]. In particular, such a situation occurs in geometrical optics of inhomogeneous media, where the SHE (also called the optical Magnus effect) has recently been predicted and examined[10–19] (not to be confused with "optical SHE" of exciton-polaritons in a semiconductor microcavity[20]).

According to theoretical predictions, a light beam propagating along a curved trajectory experiences a *polarization-dependent deflection* (SHE of light) caused by the spin-orbit interaction and solely determined by the trajectory geometry. Due to this, even a locally-isotropic inhomogeneous medium is supposed to manifest a circular birefringence of a purely topological origin[10–19]. This SHE of light in a smooth inhomogeneous medium is described by equations of motion with a "Lorentz force" from the momentum-space topological monopole, quite similarly to the SHE in semiconductors with an applied electric field[6] and to the geometrodynamics of spinning particles in external fields[10,17,18]. This offers a unique opportunity to test fundamental equations of high-energy and condensed matter physics in an optical lab setting.

Despite recent experimental efforts[21–23], direct observation of the intrinsic SHE due to the topological monopole has remained an open challenge. In high-energy physics, the observation of the SHE is far beyond the current experimental capabilities[10]. In condensed matter physics, direct measurements of the particle trajectories are impossible, and the situation is further complicated by competing extrinsic effects[21,22]. In optics, the SHE has recently been measured with a great accuracy for a sharp medium inhomogeneity and a trajectory break[23], but the Berry phase formalism is inapplicable in this non-adiabatic case. (This tiny effect, also known as the Imbert-Fedorov transverse shift[24,25], has been studied over the past 50 years but has been clarified only recently, see Refs. 14,15 and references therein.) Thus, the fundamental semiclassical equations of motion involving the topological monopole in the momentum space have not been verified yet. Here we report a unified theory and the first direct observation of the intrinsic SHE of light caused by the topological monopole underlying the adiabatic evolution of massless particles.

**Basic theory**

The geometrical optics approximation for the propagation of light in an inhomogeneous medium is a counterpart of the semiclassical formalism in quantum mechanics[26]. This short-wavelength approximation requires smallness of the parameter $\mu = \lambdabar / L << 1$, where $\lambdabar = \lambda / 2\pi$, $\lambda$ is the wavelength, and $L$ is the characteristic scale of the medium inhomogeneity. In such an approach, the propagation of light is considered as a particle-like wave-packet motion along trajectories described by the canonical formalism on the phase space $(\mathbf{p}, \mathbf{r})$. It is convenient to define the dimensionless wave momentum as $\mathbf{p} = \lambdabar_0 \mathbf{k}$ with $\mathbf{k}$ being the central wave vector and $\lambdabar_0 = c / \omega$ ($\omega$ is the wave frequency)[26]. The parameter $\lambdabar_0$, which corresponds to the wavelength in vacuum, is a counterpart of the Planck constant in the semiclassical approximation.

Unlike classical point particles, electromagnetic waves have an intrinsic property − polarization or spin − which is responsible for the intrinsic angular momentum carried by light. The spin eigenstates of photons are the right-hand (R) and left-hand (L) circular polarizations denoted by the helicity $\sigma = \pm 1$. The spin angular momentum per one photon (in units of $\hbar$) equals $\sigma \mathbf{p} / p$.

In the zero-order approximation in $\mu$ (i.e., in the "classical" $\lambdabar \to 0$ limit of wave equations), external and internal degrees of freedom of light are uncoupled from each other. In this manner, the propagation of the electromagnetic wave is independent of the polarization, and a polarization degeneracy takes place[26]. To describe the polarization evolution of light, one has to implicate the first-order approximation, which can be regarded as "semi-geometrical optics" akin to the semiclassical approximation in quantum mechanics. In this approximation, by taking the $\lambdabar$-order corrections into account, the polarization and orbital degrees of freedom become coupled with each other, which implies a *spin-orbit interaction* of photons[10,11].

This spin-orbit interaction can be described by the coupling Lagrangian (see Supplementary Information)

$$\mathcal{L}_{\text{SOI}} = -\lambdabar_0 \sigma \mathbf{A}(\mathbf{p})\dot{\mathbf{p}} \tag{1}$$

that arises under diagonalization of Maxwell equations[10,19], cf. Refs. 4,6. Hereafter, the overdot stands for the derivative with respect to the ray length $l$. The Lagrangian (1) is reminiscent of the Lagrangian $\dfrac{e}{c}\mathcal{A}(\mathbf{r})\dot{\mathbf{r}}$ of a charged particle coupled with an electromagnetic field $\mathcal{A}$, but in Eq. (1) $\mathbf{A}(\mathbf{p})$ is the Berry vector-potential (connection) that has a purely geometrical origin[27]. The same Lagrangian occurs under the evolution of relativistic spinning particles[10,17], and the Berry potential $\mathbf{A}(\mathbf{p})$ generates the topological monopole at the origin of momentum space[9,10,19]:

$$\mathbf{F} = \frac{\partial}{\partial \mathbf{p}} \times \mathbf{A} = \frac{\mathbf{p}}{p^3}. \tag{2}$$

The spin-orbit interaction (1) and Berry monopole (2) are of a dual geometro-dynamical nature. On the one hand, the fields $\mathbf{A}$ and $\mathbf{F}$ represent the connection and curvature underlying the *parallel transport* of the wave electric field[28–33]. On the other hand, the Lagrangian (1) is nothing else but the *Coriolis term* in a wave-accompanying non-inertial coordinate frame[19,34] (see Supplementary Information).

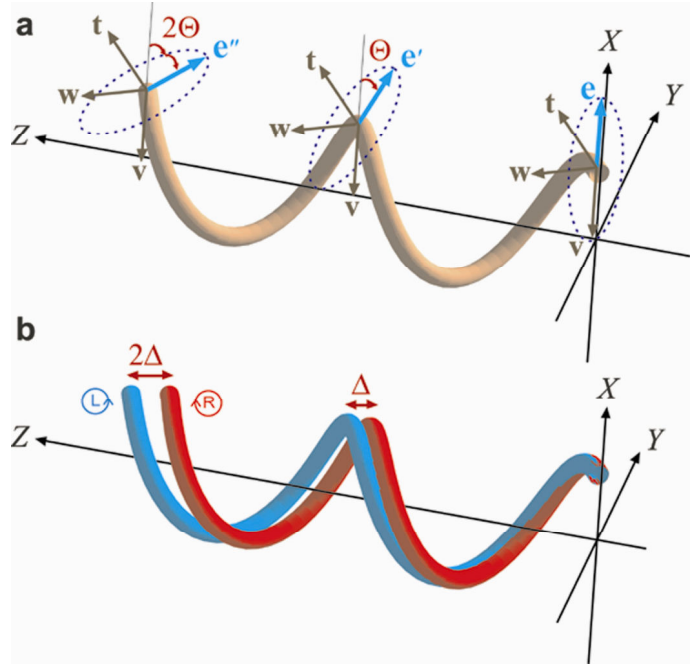

**Figure 1. Spin-orbit interaction of photons − the Berry phase and spin Hall effect − on a helical light trajectory. a**, The polarization ellipse evolves along a twisted ray trajectory, obeying the parallel transport law, see Fig. 2a. This effect is described by the Berry phase difference between the R and L polarization modes. The ray-accompanying coordinate frame is attached here to the Frenet trihedron: $(\mathbf{t}, \mathbf{v}, \mathbf{w}) = (\mathbf{t}, \mathbf{n}, \mathbf{b})$. **b**, The spin Hall effect of light. The ray trajectory is disturbed by the spin of photons, and the R and L polarized beams are deflected in the opposite directions.

There are two manifestations of the spin-orbit interaction of photons[11]. The first one is the influence of the trajectory upon the polarization. This is the Berry phase leading to the parallel transport of the wave electric field, as predicted 70 years ago by Rytov and Vladimirskii[28,29] and examined and measured in the 1980s by Ross, Tomita, Chiao, and Wu for coiled single-mode optical fibers[30−33]. The total WKB phase of the wave propagating along a ray trajectory $\ell$ is given by

$$\Phi = \lambdabar_0^{-1} \int_{\ell} \mathbf{p} \, d\mathbf{r} - \sigma \int_{\Gamma_{\ell}} \mathbf{A} \, d\mathbf{p} , \qquad (3)$$

where $\Gamma_\ell$ is the corresponding contour of the wave evolution in the **p**-space. The second term in Eq. (3) is the Berry phase due to the spin-orbit Lagrangian (1), which has opposite signs for R and L polarized waves. For elliptically polarized light, the phase difference between the R and L components determines the rotation of the polarization ellipse along the ray in accordance with the parallel transport law[28–33], Figs. 1a and 2a. The polarization state of a fully-polarized light is described by the unit three-component Stokes vector (pseudospin) $\vec{S}$ undergoing $SO(3)$ evolution on the Poincaré (Bloch) sphere. The evolution of polarization along the ray can be written as the following precession equation for the Stokes vector[19] (see Supplementary Information):

$$\dot{\vec{S}} = \vec{\Omega} \times \vec{S}, \quad \vec{\Omega} = \left(0, 0, 2\mathbf{A}\dot{\mathbf{p}}\right). \tag{4}$$

Thus, the third component of the Stokes vector is conserved upon the evolution: $S_3 = \mathrm{const}$, Fig. 2b. $S_3 \in (-1, 1)$ is the mean helicity – the expectation value of the quantum helicity $\sigma = \pm 1$ – and its conservation signifies the adiabatic evolution of photons.

It should be noted that the wave polarization is measured in a coordinate frame with basic vectors $(\mathbf{t}, \mathbf{v}, \mathbf{w})$ accompanying the ray, where $\mathbf{t} = \mathbf{p} / p$ is the tangent to the trajectory, Fig. 1a. Naturally, the Stokes parameters and gauge of the potential $\mathbf{A}$, depend on the choice of the basic vectors $(\mathbf{v}, \mathbf{w})$ at each point of the trajectory – $SO(2)$ rotations of $(\mathbf{v}, \mathbf{w})$ produce $U(1)$ gauge transformations of $\mathbf{A}$. In particular, if the coordinate frame is attached to the Frenet trihedron, $(\mathbf{t}, \mathbf{v}, \mathbf{w}) = (\mathbf{t}, \mathbf{n}, \mathbf{b})$, where $\mathbf{n}$ and $\mathbf{b}$ are the normal and binormal to the ray, the ray torsion $T^{-1}$ substitutes the quantity $-\mathbf{A}\dot{\mathbf{p}}$, so that the Berry phase equals[28–30,33] $\sigma \int_\ell T^{-1} dl$.

The second manifestation of the spin-orbit interaction of photons is the reciprocal influence of the polarization upon the trajectory of light. The coupling Lagrangian (1) brings about a polarization-dependent perturbation of the ray trajectories. As a result, the motion of the centre of gravity of a polarized wave packet is described by the equations[11–14,19]

$$\dot{\mathbf{p}} = \nabla n \,, \quad \dot{\mathbf{r}} = \frac{\mathbf{p}}{p} + \bar{\lambda}_0 S_3 \dot{\mathbf{p}} \times \mathbf{F} \,. \tag{5}$$

Here $n(\mathbf{r})$ is the refractive index of the medium (which plays the role of an external scalar potential), whereas the $\bar{\lambda}$-order term describes the "Lorentz force" caused by the topological monopole (2). Due to Eqs. (5), light beams of different polarizations propagate along slightly *different* trajectories, i.e., an effective circular birefringence occurs in an inhomogeneous (but locally isotropic!) medium, Fig. 1b. This is *the SHE of light* or *the optical Magnus effect* which is analogous to both the SHE of quantum particles[5–7,10] and the Magnus effect for quantum vortices[16,35]. The SHE of light was predicted by Liberman and Zel'dovich[11], and theoretically described by Bliokh and Bliokh[12,13] and Onoda, Murakami, and Nagaosa[14]. As was shown in Ref. 14, the topological correction in Eqs. (5) ensures the conservation of the total angular momentum of light: $\mathbf{J} = \mathbf{r} \times \mathbf{p} + \bar{\lambda}_0 S_3 \, \mathbf{p} / p = \text{const}$. This links the effect to a similar phenomenon of the Imbert−Fedorov transverse shift that appears under light reflection or refraction at sharp interfaces (where $\mu \propto \nabla n \to \infty$, and the adiabatic approximation is violated), see Refs. 14,15,23 and references therein. In terms of geometric characteristics of the ray, the topological term in the Eqs. (5) takes the form $-\bar{\lambda} S_3 R^{-1} \mathbf{b}$, where $\bar{\lambda} = \bar{\lambda}_0 / p = k^{-1}$ and $R^{-1} = |\dot{\mathbf{p}} \times \mathbf{p}| / p^2$ is the curvature of the ray.

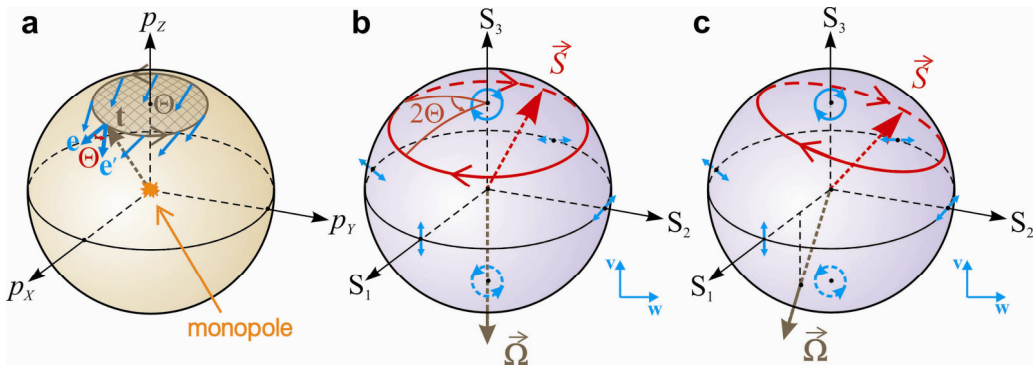

**Figure 2. Representations of the evolution of the wave polarization along a helical ray trajectory. a**, The parallel transport of the polarization vector $\mathbf{e} \perp \mathbf{t}$ on the unit $\mathbf{t}$-sphere in momentum space. For a loop trajectory, the polarization is turned on the angle $\Theta$ – the solid angle enclosed by the loop. This law is associated with the topological monopole (2) in the origin of momentum space. **b**, The same polarization evolution can equally be represented by the precession of the Stokes vector $\vec{S}$ on the Poincaré sphere about the $S_3$ axis, Eq. (4). **c**, In an anisotropic medium, the polarization evolution along the trajectory is described by a generalized precession of the Stokes vector on the Poincaré sphere, Eq. (7).

Together, Eqs. (4) and (5) form a set of coupled equations of motion for the internal and external degrees of freedom of light. Below we aim to provide an experimental verification of the effects of the spin-orbit interaction of light, particularly of the SHE described by Eqs. (5).

**Modification for a curved reflecting surface**

The experimental realization of a spiral beam as shown in Fig. 1 in a smooth gradient-index medium is a challenging problem as it requires fabrication of the appropriate smoothly inhomogeneous sample. To circumvent this, we put forth another mechanism bending the light trajectories. Namely, we consider the light grazing a curved total internal reflection surface. The multiple total internal reflections of the light beam from a concave surface at the grazing angle result in the propagation of light *along* the surface. For instance, the light beam entering a glass cylinder from the end at a tangent to the surface will propagate along the cylinder surface forming a helix similar to that in Fig. 1, see Fig. 3. The gradient-index approximation and the first Eq. (5) are inapplicable in this case since $\nabla n = \infty$ at the surface. However, the second Eq. (5) holds true (see Supplementary Information) and the SHE due to the topological monopole can be observed. As the light propagates along the surface, the Fermat principle implies that in the zero-order approximation the ray trajectory is a geodesic of the surface. Then, one can substitute the first Eq. (5) with the equation for the tangent to the geodesic. As a result we have:

$$\frac{\dot{\mathbf{p}}}{p} = \frac{\mathbf{N}}{R_N}, \ \dot{\mathbf{r}} = \frac{\mathbf{p}}{p} + \bar{\lambda}_0 S_3 \frac{\dot{\mathbf{p}} \times \mathbf{p}}{p^3}. \tag{6}$$

Here $\mathbf{N}$ is the normal to the surface, which is directed inside the dielectric, and $R_N$ is the radius of curvature of the surface cross-section including $\mathbf{t}$. It is easy to see that the normal and the curvature of the surface coincide with the normal and curvature of the ray: $\mathbf{N} = \mathbf{n}$, $R_N = R$. Equations (6) represent ray equations for the geometrical-optics light propagation along a concave reflecting surface. Here the short-wavelength approximation is assured by the smallness of the parameter $\mu' = \bar{\lambda} / L'$, $L' = \min(R, T)$, similar to the case of a bent optical fiber[30–33].

The evolution of the polarization, Eq. (4), along a totally-reflecting surface requires modification as well. The point is that the wave helicity $S_3$ is *not* conserved under the total internal reflection. Indeed, there is a phase difference that occurs between the $p$ and $s$ linearly polarized modes reflected from the surface[36]. By considering the grazing-angle limit of this phase difference, one can show that the wave undergoes an effective linear birefringence as in an *anisotropic* medium characterized by the phase difference $\sqrt{1 - n^{-2}} / R$ per unit ray length (see Supplementary Information). The anisotropy axes are naturally attached to the Frenet trihedron (the $p$ and $s$ modes are polarized along $\mathbf{n} = \mathbf{N}$ and $\mathbf{b}$, respectively), and it is convenient to write the equation for the evolution of polarization in the Frenet frame where $\mathbf{A}\dot{\mathbf{p}} \rightarrow -T^{-1}$. By introducing the phase difference between the $s$ and $p$ modes, we arrive at the modified precession equation for the Stokes vector[19]

$$\dot{\vec{S}} = \vec{\Omega} \times \vec{S}, \ \vec{\Omega} = \left( \sqrt{1 - n^{-2}} R^{-1}, 0, -2T^{-1} \right). \tag{7}$$

Thus, the linear birefringence due to the total internal reflection competes with the circular birefringence due to the Berry phase resulting in a precession of the Stokes vector about the inclined vector $\vec{\Omega}$, Fig. 2c. The helicity is not conserved there, $S_3 \neq \mathrm{const}$ (which violates conservation of the total angular momentum $\mathbf{J}$), but the polarization evolution is still smooth, so that one can regard this regime as a modified adiabatic evolution. Unlike the isotropic-medium case, Eqs. (4) and (5), the precession of the Stokes vector influences the ray deflection in Eqs. (6) via varying helicity $S_3(l)$. This causes oscillations of the light trajectory which are similar to the *zitterbewegung* of electrons with a spin-orbit interaction[19].

**Experiment**

To verify the evolution equations (6) and (7), we performed an experiment involving helical light beams propagating at a grazing angle inside a glass (BK7) cylinder. The experimental setup is shown in Fig. 3. A linearly polarized HeNe laser beam at $\lambda_0 = 633\,\mathrm{nm}$ wavelength was either right- or left-hand circularly polarized by a variable liquid crystal retarder (Meadowlark Optics). The circularly polarized beam was sent at a grazing angle into a glass cylinder, using a right angle prism fitted with an index matching gel. We used a cylinder with radius $R_0 = 8\,\mathrm{mm}$ and length $L_0 = 96\,\mathrm{mm}$ and an incident beam of 1mm width. Once inside the cylinder, the beam underwent continued internal reflections that resulted in a helical trajectory along the glass/air interface. The number of coils was adjusted by controlling the beam's angle of propagation, $\theta$, between $\mathbf{t}$ and the cylinder axis. The output Stokes parameters and the beam position were measured using a polarizer, quarter-wave plate, and CCD camera (12-bit digital-cooled, PCO Sensicam 370XL, 1280x1024 pixels) imaging the outlet face of the cylinder through a second identical right angle prism.

To calculate the output beam parameters, note that the helical ray has a constant curvature $R^{-1} = R_0^{-1}\sin^2\theta$ and torsion $T^{-1} = R_0^{-1}\sin\theta\cos\theta$. Hence, $\vec{\Omega} = \mathrm{const}$, and Eq. (6) and (7) can readily be integrated (see Supplementary Information). For R and L polarized incident beams, $\vec{S}_{\mathrm{in}}^{(R,L)} = (0,0,\pm1)$, this yields the output polarizations $\vec{S}_{\mathrm{out}}^{(R,L)}$ and the shifts of the trajectory $\delta\mathbf{r}_{\mathrm{out}}^{(R,L)}$:

$$\vec{S}_{\text{out}}^{(R,L)} = \pm\left(\omega_1\omega_3\left[1-\cos\left(\Omega l_0\right)\right], -\omega_1\sin\left(\Omega l_0\right), \left(1-\omega_3^2\right)\cos\left(\Omega l_0\right)+\omega_3^2\right), \tag{8}$$

$$\delta\mathbf{r}_{\text{out}}^{(R,L)} = \mp\lambdabar\frac{l_0}{R_0}\sin^2\theta\left[\omega_3^2+\left(1-\omega_3^2\right)\frac{\sin\left(\Omega l_0\right)}{\Omega l_0}\right]\mathbf{b}. \tag{9}$$

where $\vec{\omega}=\vec{\Omega}/\Omega$ and $l_0=L_0/\cos\theta$ is the total ray length in the cylinder. The relative output shift between the initially R and L polarized beams is $\Delta_{\text{out}}=\left[\delta\mathbf{r}_{\text{out}}^{(R)}-\delta\mathbf{r}_{\text{out}}^{(L)}\right]\mathbf{b}$. Unlike the isotropic case, $\Delta_{\text{out}}$ is a non-linear function of the ray length $l_0$ due to the influence of the Stokes vector precession. The second, oscillatory term in square brackets in Eq. (9), describes the *zitterbewegung* of the light trajectory[19].

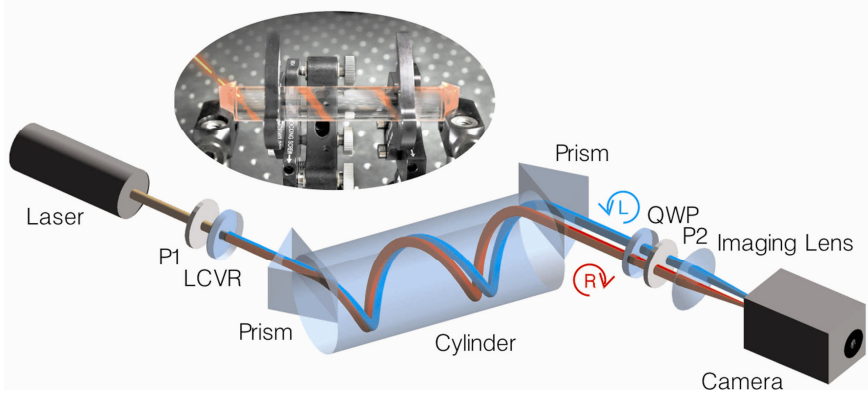

**Figure 3. Experimental setup.** A laser light beam enters the glass cylinder at a grazing angle via the input prism, coils along the cylinder surface, and leaves it via the output prism. The liquid crystal variable retarder (LCVR) is used for generating and switching between the circularly polarized modes, whereas the quarter-wave plate (QWP) and polarizer P2 are intended for measurements of the Stokes parameters. The inset shows a real picture of the spiral light beam inside the cylinder.

Figure 4 shows the output Stokes parameters $\vec{S}_{\text{out}}^{(R)}$ and the relative shift $\Delta_{\text{out}}$ as theoretically predicted by Eqs. (8) and (9) and experimentally measured at different angles of propagation. The angle of propagation is expressed by the number of turns of the helix, $m$, as $\tan\theta = 2\pi R_0\, m / L_0$. An experimental error of the Stokes parameters of 0.07 was caused by the angular tolerance of the polarization elements. The number of coils $m$ was determined to a typical accuracy of 0.2 turn. The Stokes parameters were measured using the four-measurements technique[37], while the position of the output beam was determined as a centre of mass (centroid) of the intensity distribution at the output face of the cylinder. The effects of the laser beam drift about 2μm (typical rate 0.1Hz) at the outlet face of the cylinder were minimized by taking alternating readings between the two circular polarizations at a rate of 1Hz and averaging over 40 measurements. Such technique reduced the errors that are due to the statistical noises, and, as a result, the relative shift $\Delta_{out}$ was determined to a typical accuracy of 0.3μm. Systematic errors caused by a non-normal beam incidence on the prism (the angle of incidence being up to 25 degrees) are negligible against a background of the above statistical errors.

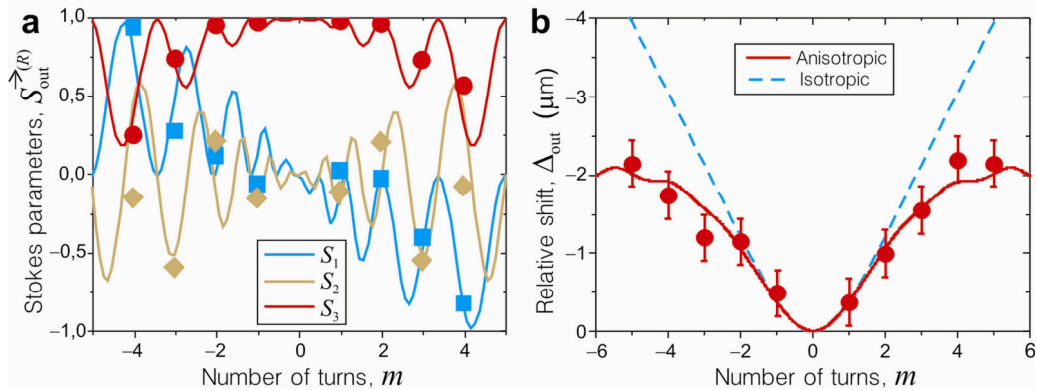

**Figure 4. Experimental measurements of the Stokes-vector precession and the spin Hall effect of light.** Theoretically calculated (curves) and experimentally measured (symbols) characteristics of the light beam as dependent on the number of turns, $m$, of the helical trajectory inside the cylinder. The negative $m$ correspond to left-hand helixes. **a**, The Stokes parameters $\vec{S}$ at the output for the R polarized incident beam, cf. Figs. 1a and 2c. The errors correspond approximately to the size of the symbols. **b**, The relative shift $\Delta$ between the output beam positions of the R and L incident beams, cf. Fig. 1b. The *zitterbewegung* variations of the beam position become noticeable at the theoretical curve for higher $|m|$. The dashed curve indicates the shift calculated in the isotropic approximation with $S_3 = \text{const}$.

The experimental measurements fully confirm all the theoretical predictions. The Stokes vector precession, the spin Hall effect, and the effect of the anisotropy on the shift $\Delta_{\text{out}}$ are clearly seen in Fig. 4.

**Conclusion**

We have presented a unified theory and a direct observation of the spin Hall effect of light and of the Stokes-vector precession in effectively inhomogeneous and anisotropic medium. Both the effects arise from the spin-orbit interaction of photons and the topological monopole in Maxwell equations. While the Berry phase and the parallel transport of polarization along the ray could be regarded as a purely *geometrical* phenomenon (it disappears in the proper parallel-transported coordinate frame), the SHE of light allows a natural *dynamical* explanation and occurs independently of the coordinate frame. Together, these two effects reveal in-depth geometro-dynamical interrelations underlying the evolution of spinning particles in external fields. Because of the close similarity and common topological roots of the SHE in optics, condensed matter, and high-energy physics, one can regard our results as indirect evidence of the intrinsic SHE in a diversity of physical systems.

In addition to the fundamental interest, the SHE has a potential application as a novel type of the particle transport. Our experiment, as well as the recent experiment Ref. 23, indicates that optical systems have an advantage over the condensed-matter systems because of the relative purity and simplicity of the optical components. Besides, the magnitude of the SHE of light can be substantially enhanced by involving higher-order angular-momentum states of light – optical vortex beams[16]. In general, introducing the spin-orbit coupling of electromagnetic waves in the contemporary photonics and nano-optics may result in the development of a promising new branch – spinoptics.

**Acknowledgements** The work by K.B. is supported by the Linkage International Grant of the Australian Research Council.

**Author information** Correspondence and requests for materials should be addressed to K.B. (k.bliokh@gmail.com).

# SUPPLEMENTARY INFORMATION

# Geometrodynamics of Spinning Light

## Konstantin Y. Bliokh, Avi Niv, Vladimir Kleiner, and Erez Hasman

Here we provide calculations underlying the theory of geometrodynamical evolution of polarized light in smooth inhomogeneous medium and along a curved reflecting surface. All the equations presented in the article are consistently derived starting from Maxwell equations.

## 1. Diagonalization of Maxwell equations, spin-orbit coupling of photons, Berry phase, and equations of motion

Maxwell equations for monochromatic electric field $\mathbf{E}$ in an inhomogeneous dissipationless dielectric medium can be written as a three-component vector equation

$$-\lambdabar_0^2 \nabla \times (\nabla \times \mathbf{E}) + n^2 \mathbf{E} = 0,$$

or,

$$(\lambdabar_0^2 \nabla^2 + n^2) \mathbf{E} - \lambdabar_0^2 \nabla (\nabla \mathbf{E}) = 0. \tag{S1}$$

where $n^2(\mathbf{r}) = \varepsilon(\mathbf{r})$ is the dielectric constant of the medium. Equation (S1) resembles the Helmholz equation, except for the last, polarization term, which mixes internal and external degrees of freedom of the wave and makes Eq. (S1) non-diagonal[S1,19]. This polarization term in Eq. (S1) guarantees that $\nabla(n^2 \mathbf{E}) = 0$ which, in turn, ensures that in a smoothly inhomogeneous medium the wave electric field remains nearly *transverse* with respect to the current momentum $\mathbf{p}$:

$$\mathbf{E} = \mathbf{E}_\perp + E_\parallel \mathbf{t}, \;\; \mathbf{E}_\perp \perp \mathbf{t}, \;\; E_\parallel \sim \mu E_\perp. \tag{S2}$$

Here $\mathbf{t} = \mathbf{p}/p$ is the unit vector tangent to the zero-order ray trajectory (S2), $E_\parallel$ is the longitudinal component of the field, and $\mathbf{E}_\perp$ is the projection of the electric field on the plane orthogonal to $\mathbf{t}$.

The wave polarization is essentially determined by the transverse field components, $\mathbf{E}_\perp$. Hence, the dimension of the problem can be reduced to 2 by *projecting* Maxwell equation (S1) onto the plane orthogonal to $\mathbf{t}$, which eliminates the longitudinal field component $E_\parallel$ from the problem. This implies a description of the wave evolution in a coordinate frame with basis vectors $(\mathbf{v}, \mathbf{w}, \mathbf{t})$ attached to the local direction of momentum, $\mathbf{t}$, Fig. 1A. Vectors $(\mathbf{v}, \mathbf{w})$ provide a natural basis of linear polarizations:

$$\mathbf{E}_\perp = E_\mathbf{v} \mathbf{v} + E_\mathbf{w} \mathbf{w}. \tag{S3}$$

However, the coordinate frame $(\mathbf{v}, \mathbf{w}, \mathbf{t})$ is *non-inertial* in the generic case; it experiences rotation as $\mathbf{t}$ varies along the ray trajectory in an inhomogeneous medium. Such rotation is described by a precession of the triad $(\mathbf{v}, \mathbf{w}, \mathbf{t})$ with some angular velocity $\boldsymbol{\Lambda}$:

$$\dot{\mathbf{v}} = \boldsymbol{\Lambda} \times \mathbf{v}, \;\; \dot{\mathbf{w}} = \boldsymbol{\Lambda} \times \mathbf{w}, \;\; \dot{\mathbf{t}} = \boldsymbol{\Lambda} \times \mathbf{t},$$

$$\boldsymbol{\Lambda} = \left( \dot{\mathbf{v}} \mathbf{w} \right) \mathbf{t} + \left( \dot{\mathbf{w}} \mathbf{t} \right) \mathbf{v} + \left( \dot{\mathbf{t}} \mathbf{v} \right) \mathbf{w} = \Lambda_\parallel \mathbf{t} + \mathbf{t} \times \dot{\mathbf{t}}, \tag{S4}$$

where $\Lambda_\parallel = \boldsymbol{\Lambda} \mathbf{t} = \dot{\mathbf{v}} \mathbf{w}$ is the longitudinal component of $\boldsymbol{\Lambda}$.

When performing a transition to the non-inertial frame $(\mathbf{v}, \mathbf{w}, \mathbf{t})$, effective inertia terms appear in Maxwell equations (S1). Similarly to classical mechanics, they can be derived via the substitution[34] $\dfrac{\partial \mathbf{E}}{\partial t} \to \dfrac{\partial \mathbf{E}}{\partial t} + \dfrac{c}{n} \boldsymbol{\Lambda} \times \mathbf{E}$, or, $\omega \to \omega + i\dfrac{c}{n} \boldsymbol{\Lambda} \times \mathbf{E}$, in Eqs. (S1). [Here the wave velocity $c/n$ occurs because we defined the angular velocity (S4) with respect to the ray length $l$ rather than time.] Neglecting higher-order terms proportional to $\Lambda^2$ and $\dot{\Lambda}$, we arrive at

$$\left( \lambdabar_0^2 \nabla^2 + n^2 \right) \mathbf{E} + 2in\lambdabar_0 \boldsymbol{\Lambda} \times \mathbf{E} + \lambdabar_0^2 \nabla \left( \nabla \mathbf{E} \right) = 0 \,. \qquad (S5)$$

The second term here is the *Coriolis term* caused by the rotation of the ray coordinate frame. It is small: $\lambdabar_0 \Lambda \sim \mu$, but should be taken into account in the first-order approximation in $\mu$.

Projecting equation (S5) onto the plane $(\mathbf{v}, \mathbf{w})$, one can show that $\left[ \lambdabar_0^2 \nabla \left( \nabla \mathbf{E} \right) \right]_{\perp} \simeq 0$ and $\left( \boldsymbol{\Lambda} \times \mathbf{E} \right)_{\perp} \simeq \Lambda_{\parallel} \left( \mathbf{t} \times \mathbf{E}_{\perp} \right)$ in the first approximation in $\mu$. Thus, the projection onto the $(\mathbf{v}, \mathbf{w})$ plane cancels the polarization term, and we have[34]

$$\left( \lambdabar_0^2 \nabla^2 + n^2 \right) \mathbf{E}_{\perp} + 2in\lambdabar_0 \Lambda_{\parallel} \left( \mathbf{t} \times \mathbf{E}_{\perp} \right) = 0 \,. \qquad (S6)$$

Equation (S6) is a two-component vector equation which becomes diagonal in the basis of *circular* polarizations. Substituting the field as a superposition of right- and left-hand modes,

$$\mathbf{E}_{\perp} = E^{+} \boldsymbol{\xi} + E^{-} \boldsymbol{\xi}^{*} \,, \; \boldsymbol{\xi} = \dfrac{\mathbf{v} + i\mathbf{w}}{\sqrt{2}} \,, \; E^{\pm} = E_{\mathbf{v}} \mp i E_{\mathbf{w}} \,, \qquad (S7)$$

we obtain

$$\left( \lambdabar_0^2 \nabla^2 + n^2 \right) E^{\sigma} + 2n\lambdabar_0 \sigma \Lambda_{\parallel} E^{\sigma} = 0 \,. \qquad (S8)$$

Hereafter $\sigma = \pm 1$ denotes the wave *helicity* indicating the two spin states of photons. Equation (S8) shows that, in the first approximation of geometrical optics, these two states evolve independently, and the zero-order polarization degeneracy is removed by the Coriolis term.

Performing substitution $-i\lambdabar_0 \nabla \rightarrow \mathbf{p}$ in Eq. (S8), we find a characteristic equation which gives the wave Hamiltonian

$$\mathcal{H} = p - n - \lambdabar_0 \boldsymbol{\sigma} \boldsymbol{\Lambda} = 0 \,. \qquad (S9)$$

Here we simplified the characteristic equation in the first approximation in $\mu$ and introduced the spin angular momentum per photon (in units of $\hbar$), $\boldsymbol{\sigma} = \sigma \mathbf{t}$, so that $\boldsymbol{\sigma}\boldsymbol{\Lambda} = \sigma \Lambda_\parallel$. As compared to the traditional geometrical optics Hamiltonian[26], Eq. (S9) contains an additional spin term which is equivalent to the Coriolis term of spinning particles in a rotating frame[S2,S3]. The Lagrangian corresponding to the Hamiltonian (S9) takes the form

$$\mathcal{L} = \mathbf{p}\dot{\mathbf{r}} - p + n + \lambdabar_0 \boldsymbol{\sigma} \boldsymbol{\Lambda} \equiv \mathcal{L}_0 + \mathcal{L}_{SOI} \,. \qquad (S10)$$

where $\mathcal{L}_0 = n - p + \mathbf{p}\dot{\mathbf{r}}$ is the scalar-approximation Lagrangian, whereas $\mathcal{L}_{SOI} = \lambdabar_0 \boldsymbol{\sigma} \boldsymbol{\Lambda}$ is the Lagrangian describing the *spin-orbit coupling* of photons. It should be noted that the spin-orbit term in the Lagrangian had been known for spinning particles before the Berry phase discovery[S4,S5].

In order to represent the spin-orbit Lagrangian in the Berry-phase form, we notice that the co-moving coordinate frame $(\mathbf{v}, \mathbf{w}, \mathbf{t})$ is attached to the direction of the wave momentum $\mathbf{p}$, and the polarization evolution of the wave is essentially *momentum-dependent* (rotations of the ray coordinate frame are independent of the particular space coordinates, $\mathbf{r}$ ). Therefore, we can parametrize the basis vectors of the ray coordinate frame as:

$$\mathbf{t} = \mathbf{t}(\mathbf{p}), \ \mathbf{v} = \mathbf{v}(\mathbf{p}), \ \mathbf{w} = \mathbf{w}(\mathbf{p}). \tag{S11}$$

The transition from the $l$-parametrization to the $\mathbf{p}$-parametrization is performed via the substitution $\dfrac{d}{dl} \rightarrow \dfrac{d\mathbf{p}}{dl}\dfrac{\partial}{\partial \mathbf{p}}$, and the spin-orbit Lagrangian (S10) takes the form of Eq. (1):

$$\mathcal{L}_{\text{SOI}} = -\bar{\lambda}_0 \sigma \mathbf{A}(\mathbf{p}) \dot{\mathbf{p}}, \tag{S12}$$

where

$$A_i = \mathbf{v}\frac{\partial \mathbf{w}}{\partial p_i} = -i\xi^* \frac{\partial \xi}{\partial p_i} \tag{S13}$$

is the *Berry connection* or the *Berry gauge field*.

The Berry connection relates the wave polarization $\mathbf{e} = \mathbf{E}_\perp / E_\perp$ in the neighboring points $\mathbf{p}$ and $\mathbf{p} + d\mathbf{p}$ of momentum space. Note that the polarization evolution depends only on the *direction* of momentum, $\mathbf{t} = \mathbf{p}/p$. Therefore, the evolution in the $\mathbf{p}$ space can be projected onto the unit sphere $\mathrm{S}^2 = \{\mathbf{t}\}$ in momentum space. In this manner, the polarization vector $\mathbf{e}$ is tangent to this sphere, and the Berry connection determines the natural *parallel transport* of $\mathbf{e}$ over the $\mathrm{S}^2$ sphere[2,29–33,S6,S7], Fig. 2A.

The curvature tensor corresponding to the connection (S13) is $F_{ij} = \dfrac{\partial A_j}{\partial p_i} - \dfrac{\partial A_i}{\partial p_j}$ which yields

$$F_{ij} = \frac{\partial \mathbf{v}}{\partial p_i}\frac{\partial \mathbf{w}}{\partial p_j} - \frac{\partial \mathbf{v}}{\partial p_j}\frac{\partial \mathbf{w}}{\partial p_i} = -i\left(\frac{\partial \xi^*}{\partial p_i}\frac{\partial \xi}{\partial p_j} - \frac{\partial \xi^*}{\partial p_j}\frac{\partial \xi}{\partial p_i}\right). \tag{S14}$$

This the *Berry curvature* or the *Berry field strength*. It is an antisymmetric tensor which can be characterized by the dual vector $\mathbf{F} = \dfrac{\partial}{\partial \mathbf{p}} \times \mathbf{A}$, $F_{ij} = \varepsilon_{ijk} F_k$. For electromagnetic waves the Berry curvature takes the form of the "magnetic monopole" Eq. (2)[2,9,10,31,S1]:

$$\mathbf{F} = \frac{\mathbf{p}}{p^3}. \tag{S15}$$

Hence, on the surface of the unit $\mathbf{t}$-sphere (i.e., at $p = 1$) it equals $\mathbf{F} = \mathbf{t}$ indicating the unit Gaussian curvature of the sphere surface.

Note that gauge properties of the potential $\mathbf{A}$ and field $\mathbf{F}$ are directly related to the choice of the co-moving frame $(\mathbf{v}, \mathbf{w}, \mathbf{t})$, which is determined up to an arbitrary rotation about $\mathbf{t}$. Such a local rotation of the coordinate frame on an angle $\alpha = \alpha(\mathbf{p})$, induces the gauge transformation of the basic vector of circular polarizations, $\boldsymbol{\xi}$, Eq. (S7):

$$\boldsymbol{\xi} \to \exp(-i\alpha)\boldsymbol{\xi}, \tag{S16}$$

i.e. $\mathrm{SO}(2)$ rotation of $(\mathbf{v}, \mathbf{w})$ is equivalent to $\mathrm{U}(1)$ gauge transformation of $\boldsymbol{\xi}$ (see M. V. Berry in Ref. 2). In turn, the gauge transformation (S16) generates the transformation of the Berry connection (S13) but does not influence the Berry curvature (S14):

$$\mathbf{A} \to \mathbf{A} - \frac{\partial \alpha}{\partial \mathbf{p}}, \; \mathbf{F} \to \mathbf{F}. \tag{S17}$$

Therefore, all the physical quantities which are independent of the coordinate frame (e.g., ray trajectories), should depend on the Berry curvature $\mathbf{F}$ rather than on the gauge-dependent connection $\mathbf{A}$.

The total phase of the wave, Eq. (3), follows directly from the Lagrangian (S10) with the spin-orbit part in the form of Eq. (S12):

$$\Phi = \mathchar'26\mkern-9mu\lambda_0^{-1} \int_\ell \mathcal{L} \, dl = \mathchar'26\mkern-9mu\lambda_0^{-1} \int_\ell \mathbf{p} \, d\mathbf{r} - \sigma \int_{\Gamma_\ell} \mathbf{A} \, d\mathbf{p} \, , \qquad (S18)$$

where we took into account that $\mathcal{H} = 0$, Eq. (S9). Different representations of the polarization evolution stemming from the Berry phase in (S18) are considered below. The Euler-Lagrange equations with the Lagrangian (S10) and (S12) written in the form $\mathcal{L} = \mathcal{L}(\mathbf{p}, \dot{\mathbf{p}}, \mathbf{r}, \dot{\mathbf{r}})$ and varying independently with respect to $\mathbf{p}$ and $\mathbf{r}$ result in the equations of motion

$$\dot{\mathbf{p}} = \nabla n \, , \; \dot{\mathbf{r}} = \frac{\mathbf{p}}{p} + \mathchar'26\mkern-9mu\lambda_0 \sigma \dot{\mathbf{p}} \times \mathbf{F} = \frac{\mathbf{p}}{p} + \mathchar'26\mkern-9mu\lambda_0 \sigma \frac{\dot{\mathbf{p}} \times \mathbf{p}}{p^3} \, . \qquad (S19)$$

They describe the split ray trajectories of the two circularly polarized eigenmodes $\sigma = \pm 1$ of the problem. The ray equations (5) are obtained from (S19) via substitution $\sigma \rightarrow S_3$; they describe the center-of-gravity position of the beam with an arbitrary polarization.

## 2. Evolution of the polarization of light in different representations

The polarization state of the wave can be described by the unit complex two-component Jones vector in the basis of circular polarization:

$$|\psi\rangle = \begin{pmatrix} e^+ \\ e^- \end{pmatrix}, \; \langle \psi | \psi \rangle = 1 \qquad (S20)$$

where $e^\pm = E^\pm / E_\perp$ are the normalized amplitudes of the two modes. The Berry phase, Eqs. (3) and (S18),

$$\Phi_B = -\sigma \int_{\Gamma_\ell} \mathbf{A}\, d\mathbf{p} , \qquad (S21)$$

acquired by the circularly polarized components with the opposite signs, indicates the following evolution of the Jones vector:

$$\left| \psi(l) \right\rangle = \begin{pmatrix} \exp(-i\Phi_B) & 0 \\ 0 & \exp(+i\Phi_B) \end{pmatrix} \left| \psi(0) \right\rangle . \qquad (S22)$$

It is easy to see that, in the generic case of an elliptical polarization, this equation describes the turn of the polarization ellipse on angle $\Phi_B$ with its eccentricity conserved, Fig. 1A.

The differential form of Eq. (S22) is:

$$\left| \dot{\psi} \right\rangle = -i \left( \mathbf{A}\dot{\mathbf{p}} \right) \hat{\sigma}_3 \left| \psi \right\rangle , \qquad (S23)$$

where we use the Pauli matrices $\hat{\vec{\sigma}} = \left( \hat{\sigma}_1, \hat{\sigma}_2, \hat{\sigma}_3 \right)$. Equation (S23) was obtained in Refs. 14,19 and is similar to the equation for the polarization evolution in other spin systems[S8,S9]. Eq. (S23) describes a local *inertia* of the electric field which remains *locally* non-rotating about the ray. Indeed, one can show that Eq. (S23) is equivalent to the equation for the unit electric field vector $\mathbf{e} = \mathbf{E}_\perp / E_\perp$ :

$$\dot{\mathbf{e}} = -\left( \mathbf{e}\dot{\mathbf{t}} \right) \mathbf{t} . \qquad (S24)$$

This is a well-know equation for the parallel transport of vector $\mathbf{e}$ along the ray[2]. According to Eq. (S24), $\mathbf{e}$ does not experience local rotation about $\mathbf{t}$ .

As is known, there is an alternative formalism describing the polarization state of light, namely, the Stokes parameters representing the polarization state on the Poincaré sphere. This formalism is quite similar to the quantum mechanical Bloch-sphere representation. Indeed, the Stokes vector is a three-component unit vector defined as

$$\vec{S} = \langle \psi | \hat{\vec{\sigma}} | \psi \rangle, \ \vec{S}^2 = 1. \tag{S25}$$

The north and south poles of the Poincaré sphere ( $S_3 = \pm 1$ ) represent the right- and left-hand circular polarizations, whereas the equator ( $S_3 = 0$ ) represents the linear polarizations, Fig. 2B. By differentiating expression (S25) and using Eq. (S23), we find that the Stokes vector obeys the following precession equation[19]:

$$\dot{\vec{S}} = \vec{\Omega} \times \vec{S}, \ \vec{\Omega} = 2(\mathbf{A}\dot{\mathbf{p}})\vec{u}_3, \tag{S26}$$

where $\vec{u}_3$ is the basis vector of the $S_3$ axis. Thus, the Stokes vector precesses about the $S_3$ axis on the Poincaré sphere with the angular velocity $\Omega = 2\mathbf{A}\dot{\mathbf{p}}$, Fig. 2B. In the ray-accompanying coordinate frame attached to the Frenet trihedron, $(\mathbf{t}, \mathbf{v}, \mathbf{w}) = (\mathbf{t}, \mathbf{n}, \mathbf{b})$, we have[S10] $\mathbf{A}\dot{\mathbf{p}} \to -T^{-1}$. Therefore, one period of the helical ray in Fig. 1A causes azymuthal rotation of the Stokes vector on the angle[33]

$$2\Phi_B = -2\int T^{-1} dl = -4\pi + 2\Theta \tag{S27}$$

on the Poincaré sphere, Fig. 2B, where $\Theta$ is the solid angle enclosed by the trajectory of the $\mathbf{t}$ vector on the unit sphere in momentum space, Fig. 2A. The factor of 2 occurs in the evolution of the Stokes vector because a complete $2\pi$ turn of the polarization ellipse in the real space corresponds to a $4\pi$ double-turn on the Poincaré sphere.

The above two forms of the equation of the polarization evolution, Eqs. (S23) and (S26), i.e., the Jones and Stokes representations, are the optical counterparts of the Schrödinger and Heisenberg pictures of spin-1/2 evolution in quantum mechanics[19,S11,S12]. Of course, the spin of a photon is 1, but the Stokes vector is rather a *pseudo*-spin in the problem with two polarization modes.

# 3. Modified theory for the light propagation along a reflecting surface

Let us consider a number of total internal reflections of a light beam at a nearly grazing angle from a concave dielectric interface, Fig. S1. In the limit of zero angle $\alpha \to 0$ the distance between two successive reflections $\Delta l \to 0$, and the beam propagates along the smooth surface. Using the geometrical features of the reflection, which is symmetric with respect to the normal to the surface, $\mathbf{N}$, lying in the plane of propagation, we conclude that the normal to the sliding ray coincides with the normal to the surface: $\mathbf{n} = \mathbf{N}$. As a consequence, the radii of curvature of the ray, $R$, and the surface cross-section including the ray tangent $\mathbf{t}$, $R_N$, coincide: $R_N = R$. The Frenet-Serret formula for the evolution of the tangent $\mathbf{t}$ reads $\dot{\mathbf{t}} = \dfrac{\mathbf{n}}{R}$. Substituting here parameters of the ray with the parameters of the surface and using $\mathbf{t} = \mathbf{p}/p$ with $p = \mathrm{const}$ in the homogeneous dielectric medium, we arrive at

$$\frac{\dot{\mathbf{p}}}{p} = \frac{\mathbf{N}}{R_N} \,. \tag{S28}$$

This is the first modified equation of motion Eq. (6).

In order to show that the second Eq. (6) is valid, we consider the Imbert-Fedorov transverse shift $\Delta \mathbf{r}_{IF}$ at a single total internal reflection (see Refs. 14,15 and references therein), Fig. S1. The transverse shift out of the propagation plane is directed along the binormal $\mathbf{b}$ to the ray. Using the formula obtained in Ref. 15 and expressing it in terms of Stokes parameters, the transverse shift at a total internal reflection can be written as

$$\Delta \mathbf{r}_{IF} = -\hbar \tan \alpha \left[ S_3 \left( 1 + \mathrm{Re}\, \frac{\rho_\perp}{\rho_\parallel} \right) + S_2 \, \mathrm{Im}\, \frac{\rho_\perp}{\rho_\parallel} \right] \mathbf{b} \,. \tag{S29}$$

Here $\rho_{\parallel}$ and $\rho_{\perp}$ are the Fresnel reflection coefficients for the waves linearly polarized, respectively, along $\mathbf{n}$ and $\mathbf{b}$ [36]. In the limit $\alpha \to 0$ we have

$$\rho_{\perp} / \rho_{\parallel} \approx 1 + 2i\alpha\sqrt{1 - n^{-2}} \tag{S30}$$

and

$$\Delta \mathbf{r}_{IF} \approx -2\hbar\alpha S_3 \mathbf{b} . \tag{S31}$$

The ray length between two successive reflections equals $\Delta l \approx 2\alpha R$. Hence, in the limit $\alpha \to 0$ the Imbert-Fedorov shift per unit ray length, $\dfrac{\Delta \mathbf{r}_{IF}}{\Delta l} \to \delta\dot{\mathbf{r}}$, leads to the differential equation

$$\delta\dot{\mathbf{r}} = -\hbar S_3 R^{-1}\mathbf{b} . \tag{S32}$$

Taking into account that $R^{-1}\mathbf{b} = -\mathbf{p} \times \dot{\mathbf{p}} / p^2$ and $\hbar = \hbar_0 / p$, Eq. (S32) gives precisely the topological term in the second Eq. (6).

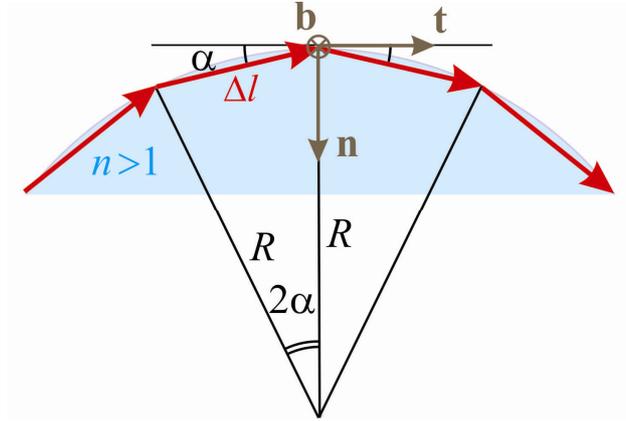

**Fig. S1.** Geometry of the successive total internal reflections from an element of a concave surface at a nearly grazing angle.

Finally, we aim to derive the modified equation for the polarization evolution along the surface. Equation (S30) implies that the $p$ and $s$ linearly polarized modes acquire the phase difference as the wave is reflected from the surface. This phase difference equals

$$\Delta\Phi = \Phi_{\mathbf{n}} - \Phi_{\mathbf{b}} \approx -2\alpha\sqrt{1-n^{-2}} \, , \qquad (S33)$$

where $\Phi_{\mathbf{n}}$ and $\Phi_{\mathbf{b}}$ are the phase acquired by the $p$ and $s$ modes (which are polarized along $\mathbf{n}$ and $\mathbf{b}$. Thus, in the limit $\alpha \to 0$, the phase difference per unit ray length yields $\dfrac{\Delta\Phi}{\Delta l} \to \delta\dot{\Phi} = -R^{-1}\sqrt{1-n^{-2}}$. This phase difference is described by the following differential equation for the Jones vector in the basis of linear polarizations:

$$\frac{d}{dl}\begin{pmatrix} E_{\mathbf{n}} \\ E_{\mathbf{b}} \end{pmatrix} = -i\frac{R^{-1}\sqrt{1-n^{-2}}}{2}\hat{\sigma}_3\begin{pmatrix} E_{\mathbf{n}} \\ E_{\mathbf{b}} \end{pmatrix}, \qquad (S34)$$

where $E_{\mathbf{n}}$ and $E_{\mathbf{b}}$ are the electric field projections on $\mathbf{n}$ and $\mathbf{b}$. By performing transformation to the basis of circularly polarized modes $E^{\pm} = E_{\mathbf{n}} \mp iE_{\mathbf{b}}$ (cf. Eqs. (S3), (S7), and (S20)) and adding the Berry phase term, Eq. (S23), we arrive at the following equation for the polarization evolution:

$$|\dot{\psi}\rangle = -i\left[ -T^{-1}\hat{\sigma}_3 + \frac{R^{-1}\sqrt{1-n^{-2}}}{2}\hat{\sigma}_1 \right]|\psi\rangle, \qquad (S35)$$

where we took into account that in the Frenet coordinate frame the Berry term $-\mathbf{A}\dot{\mathbf{p}}$ takes the form of the ray torsion $T^{-1}$.

Similarly to Eqs. (S23), (S25), and (S26), the transition to the Stokes-vector representation leads to the equation

$$\dot{\vec{S}} = \vec{\Omega} \times \vec{S}, \quad \vec{\Omega} = \left( \sqrt{1-n^{-2}}R^{-1}, 0, -2T^{-1} \right), \qquad (S36)$$

which is the equation of motion (7). Eqs. (S35) and (S36) are of the form of the polarization evolution equations in an anisotropic medium with a linear birefringence due to the curvature term and a circular birefringence due to the torsion (Berry phase) term (see, e.g., Refs. S11−S16). In the Jones representation, the polarization evolution equation for a curved rays in an anisotropic medium was obtained by Kravtsov[S17,S18], while in the Stokes vector representation it was recently derived in Refs. 19,S12,S19. It should be noted that unlike the quadratic effect of the ray curvature in an isotropic medium[30,33,S20,S21], the effective linear birefringence in Eqs. (S35) and (S36) is of the *first* order in the curvature $R^{-1}$.

# 4. Calculations for a helical ray trajectory inside a dielectric cylinder

Here we integrate the equations of motion (6) and (7) for a helical ray trajectory with a constant curvature and torsion. The solution of Eq. (7) for the output Stokes vector $\vec{S}(l)$ is expressed by the Rodrigues formula[S22] for the rotation of the initial Stokes vector $\vec{S}_{\text{in}}$ on the angle $\Omega l$ about $\vec{\omega} = \vec{\Omega}/\Omega$ :

$$\vec{S}_{\text{out}} = \vec{S}_{\text{in}} \cos(\Omega l) + (\vec{\omega} \times \vec{S}_{\text{in}}) \sin(\Omega l) + (\vec{\omega} \vec{S}_{\text{in}}) \vec{\omega} \left[ 1 - \cos(\Omega l) \right]. \qquad (S37)$$

For a right- and left-hand circularly polarized incident wave, $\vec{S}_{\text{in}}^{(R,L)} = (0, 0, \pm 1)$, Eq. (S37) yields Eq. (8):

$$\vec{S}^{(R,L)} = \pm \left( \omega_1 \omega_3 \left[ 1 - \cos(\Omega l) \right], -\omega_1 \sin(\Omega l), \left( 1 - \omega_3^2 \right) \cos(\Omega l) + \omega_3^2 \right). \qquad (S38)$$

The trajectory displacement $\delta \mathbf{r}$ is obtained by integration of Eq. (S32).

$$\delta \mathbf{r} = -\hbar R^{-1} \mathbf{b} \int_0^l S_3 \, dl = -\hbar R^{-1} l \mathbf{b} \langle S_3 \rangle \,. \tag{S39}$$

Here $\langle S_3 \rangle = \dfrac{1}{l} \displaystyle\int_0^l S_3 \, dl$ is the averaged value of the wave helicity on the ray trajectory.

Substituting Eq. (S38), we obtain

$$\left\langle S_3^{(R,L)} \right\rangle = \pm \left[ \omega_3^2 + \left( 1 - \omega_3^2 \right) \frac{\sin\left(\Omega l\right)}{\Omega l} \right]. \tag{S40}$$

Equation (S39) with (S40) gives Eq. (9) for the trajectory displacement. For a helical ray with radius $R_0$ and angle of propagation $\theta$, the ray curvature and torsion are equal, respectively, to $R^{-1} = R_0^{-1} \sin^2 \theta$ and $T^{-1} = R_0^{-1} \sin\theta\cos\theta$, which should be substituted in the above equations (S36)–(S40).

## Supplementary references